\documentclass[aps,pra,draft,showpacs,nofootinbib,preprint]{revtex4-1}

\usepackage{xcolor}
\usepackage{braket}
\usepackage{slashed}
\usepackage{mathtools}
\usepackage{nicefrac}
\usepackage{amssymb}
\usepackage{mathrsfs}
\usepackage{IEEEtrantools}
\definecolor{BrickRed}{cmyk}{0, .89, .94, .28} 

\newcommand{\be}{\begin{equation}}
\newcommand{\ee}{\end{equation}}
\newcommand{\bea}{\begin{IEEEeqnarray}{rCl}}
\newcommand{\eea}{\end{IEEEeqnarray}}

\newcommand{\ba}{\begin{array}}
\newcommand{\ea}{\end{array}}

\definecolor{lightGray}{RGB}{220,220,220}

\definecolor{MyDarkBlue}{rgb}{0,0.08,0.45}

\definecolor{MyDarkGray}{RGB}{140,140,140}

\newcommand{\bes}{\begin{equation}\begin{split}}
\newcommand{\ees}{\end{split}\end{equation}}

\definecolor{lightGray}{RGB}{220,220,220}
\definecolor{MyDarkGray}{RGB}{140,140,140}

\definecolor{MyDarkBlue}{rgb}{0,0.08,0.45}

\newlength{\eqboxstorage}

\newcommand{\ed}{\end{document}}

\newcommand{\bl}{\begin{list}{$\circ$}{}}
\newcommand{\el}{\end{list}}

\newcommand{\psib}{\bar{\psi}}

\definecolor{cadmiumgreen}{rgb}{0.0, 0.42, 0.24}

\usepackage{graphicx}% Include figure files
\usepackage{dcolumn}% Align table columns on decimal point
\usepackage{bm}% bold math
\usepackage{xcolor}
\newcommand{\vext}{v_{\rm ext}}
\newcommand{\vhxc}{v_{\rm Hxc}}
\newcommand{\Op}{\hat{\mathcal{O}}}

\newtheorem{theorem}{Theorem}[section]

\begin{document}

\title{On the Kohn-Sham Approach to Time-Dependent Problems in a Density-Functional Framework}

\author{Walter Tarantino}
\email{walter.tarantino@polytechnique.edu}

\affiliation{Laboratoire des Solides Irradi\'es, \'Ecole Polytechnique, CNRS, CEA/DSM, 91128 Palaiseau, France.}

%\author{E. K. U. Gross}
%\affiliation{Max-Planck-Institut f\"{u}r Mikrostrukturphysik, Weinberg 2, D-06120 Halle, Germany.}

\date{May 2015} % \date{\today}

%%%%%%%%%%%%%%%%%%%%%%%%%%%%%%%%%%%%%%%%%%%%%%%%%%%%%%%%%%%%%%%%%%%%%%%%%%%%%%%%%%%%%%%
\begin{abstract}
Predictivity of the Kohn-Sham approach to dynamical problems, 
when regarded as an initial value problem in a time-dependent density functional framework,
is analysed for a class of models for which the argument devised 
in the work of Maitra \textit{et al.} \cite{neepacausality}
for the standard electronic many-body problem does not apply. 
The original argument is here extended and revised. As a result, predictivity 
for this class of problems seems possible only at the price of introducing 
extra unknown functionals in the corresponding Kohn-Sham equation.
Furthermore, the same argument, when applied to original
electronic problem, suggests that the Hartree-exchange-correlation potential
is not unambiguously identified by the contemporary and past densities and initial states,
but also requires knowledge of the divergence of the contemporary Kohn-Sham current.
\end{abstract}
%%%%%%%%%%%%%%%%%%%%%%%%%%%%%%%%%%%%%%%%%%%%%%%%%%%%%%%%%%%%%%%%%%%%%%%%%%%%%%%%%%%%%%%
\pacs{
31.15.ee, %Time-dependent density functional theory
}
%\keywords{}
%%%%%%%%%%%%%%%%%%%%%%%%%%%%%%%%%%%%%%%%%%%%%%%%%%%%%%%%%%%%%%%%%%%%%%%%%%%%%%%%%%%%%%%

\maketitle
\section*{Introduction}

Time-dependent density functional theory (TDDFT) in the Kohn-Sham approach allows,
in principle, to calculate the charge density of the electronic many-body problem
by solving the equation of motion of a system of non-interacting electrons
\cite{rungegross,vL1999}.
The same approach has been adopted for other many-body Hamiltonians and
expectation value of local operators, usually refer to as `densities'.
Predictivity of the original approach was questioned in \cite{sch}
and immediately clarified in \cite{neepacausality}.

This work is intended as an extension of the argument of \cite{neepacausality}
and moves a step forth towards a complete description of the mathematical foundations of the
Kohn-Sham approach within TDDFT as an initial-value problem.

In the following I will improve the argument of \cite{neepacausality}
for the standard TDDFT case and explain how it does not hold for some other class 
of many-body problems, including some models of physical interest.
I will discuss the consequences of the failure of such an argument,
showing how the issue of `predictivity'
can be discussed in terms of existence of a unique solution 
of the equations of motion of the Kohn-Sham system.
I shall then characterize such a class of problems and, finally, 
explain how in some cases the approach can still be made `predictive', as long as 
new (unknown) density-functionals are introduced in the relevant equations.

In Section \ref{Sprocedure} the Kohn-Sham approach within the framework of 
Density-Functional Theory for time-dependent problems is recalled;
in Section \ref{Shubbard} a simple model (Hubbard dimer) for which such a procedure is problematic is presented; in Section \ref{Sprocedure2}
a tool for understanding whether a model suffers 
of the same problem is devised and a simple modification of the original recipe 
that fixes the problem for the Hubbard case is generalized.

%%%%%%%%%%%%%%%%%%%%%%%%%%%%%%%%%%%%%%%%%%%%%%%%%%%%%%%%%%%%%%%%%%%%%%%%%%%%%%%%%%%%%%%%
%%%%        %%%%        %%%%        %%%%        %%%%        %%%%        %%%%        %%%%        %%%%        %%%%        %%%%        
%%%%%%%%%%%%%%%%%%%%%%%%%%%%%%%%%%%%%%%%%%%%%%%%%%%%%%%%%%%%%%%%%%%%%%%%%%%%%%%%%%%%%%%%

\section{The Density-Functional Time-Dependent Kohn-Sham Approach (1 of 2)}\label{Sprocedure}

TDDFT in its KS declination represents an alternative to the time-dependent Schr\"odinger equation
for describing the evolution of the charge density of a system of interacting electrons \cite{ullrich}.
In fact such a reformulation of the electronic problem can be generalized and used for other
time-dependent problems as well.
The common way to perform such a generalization and apply it to other models
is the object of this section.

We start with a generic time-evolution quantum problem, formulated in second quantization
and characterized by the equation of motion
\be\label{MBeom}
i\partial_t \ket{\Psi(t)}=\left( \hat{T}+\hat{W}+\hat{V}(t)\right)\ket{\Psi(t)},
\ee
which includes a kinetic term $\hat{T}$, an interaction term $\hat{W}$ and the action
of an external classical field via $\hat{V}(t)$,
and by a given state $\ket{\Psi_0}$ such that
\be\label{MBiv}
\ket{\Psi(t=0)}=\ket{\Psi_0}.
\ee
I assume that the first order differential equation (\ref{MBeom}) has one and only one solution
$\ket{\Psi(t)}$ satisfying (\ref{MBiv}).
I shall refer to this as the `time-dependent many-body problem',
or, simply, `the many-body (MB) problem'.

If we are not interested in finding the state vector that solves the problem, but only in the expectation values 
of a finite number of local operators (from now on `densities'),
we can attempt to approach the problem via the Kohn-Sham method,
which consists in solving an auxiliary differential equation that is only linear in the field
and characterized by a set of effective potentials such that the densities evaluated in 
such a system match their value in the original problem.
Let us write $\hat{V}(t)=\sum \vext(t) \hat{n}$, where the sum is a shorthand notation 
for a integration/sum over all residual degrees of freedom 
(space/lattice points, vectorial/tensorial/spin components...)
and $\hat{n}$ is a local operator bilinear in the field.
For sake of argument, I consider the case in which we are interested only in the expectation value of $\hat{n}$,
namely $\{\braket{\Psi (t)|\hat{n}|\Psi(t)}\;|\;t\in [0,+\infty)\}$.
The Konh-Sham system for this problem is then identified by the equation of motion
\be\label{KSeom}
i\partial_t \ket{\Phi(t)}=\left( \hat{T}+\hat{V}_s(t)\right)\ket{\Phi(t)}
\ee
with $\hat{V}_s(t) \equiv \sum v_s(t) \hat{n}$, and a certain initial state 
\be\label{KSiv}
\ket{\Phi(t=0)}=\ket{\Phi_0}
\ee
such that $\braket{\Psi (t)|\hat{n}|\Psi(t)}=\braket{\Phi (t)|\hat{n}|\Phi(t)}$ for all $t\in [0,+\infty)$.
The existence of such a Kohn-Sham system is in general not known 
and it becomes crucial to prove it even under restrictive assumptions.
To proceed with my argument, 
I assume that a theorem similar to the van Leeuwen-Runge-Gross theorem \cite{rungegross,vL1999}
can in fact be proven for the model under exam. Such a theorem ensures existence and uniqueness of the system,
under specific conditions, and allows to rewrite the problem in a density-functional framework.
It can be stated in the following terms.

Given $f(t)$ a generic function of time, we distinguish the set $\{f(t)|\;t\in[0,+\infty)\}$ denoted by simply $f$,
from the value of the function \textit{at the specific time $t$} denoted by $f(t)$. 
A \textit{set of functions} is then denoted by $\mathfrak{F}\equiv\{f,g,...\}$.
The expectation value of an observable $\Op$ is denoted by $\mathcal{O}_{MB}(t)$,
if sandwiched by $\ket{\Psi(t)}$, or by $\mathcal{O}_{KS}(t)$, if sandwiched by $\ket{\Phi(t)}$.
Moreover, $n(t)\equiv n_{MB}(t)=n_{KS}(t)$.
Then, given a certain set of external potentials $\{\vext,\vext',\vext'',...\}\equiv \mathfrak{V_e}$,
we denote by $\mathfrak{N}\equiv\{n,n',n'',...\}$ the set of densities obtained 
by solving the problem (\ref{MBeom}-\ref{MBiv}) with any element of $\mathfrak{V_e}$
(giving rise to the set of `v-representable densities').
The first part of the theorem is then stated as following:
\begin{theorem}
There is a one-to-one correspondence between elements of $\mathfrak{V_e}$
and  elements of $\mathfrak{N}$, up to gauge transformations.
\end{theorem}
Moreover, given the set $\mathfrak{M}$ of densities obtained by solving (\ref{KSeom}-\ref{KSiv})
by means of some effective potentials,
the second part of the theorem reads
\begin{theorem}
There exists a set $\mathfrak{V_s}\equiv \{v_s,v_s',v_s'',...\}$
such that $\mathfrak{M}=\mathfrak{N}$ and, moreover, elements of $\mathfrak{V_s}$
are in one-to-one correspondence with $\mathfrak{N}$, up to gauge transformations.
\end{theorem}
From now on I shall refer to these two statements as to `vLRGl'
(van Leeuwen-Runge-Gross-like theorem). For the electronic many-body problem the theorem has been 
proved under some restrictive assumptions \cite{rungegross,vL1999,fixed},
which for the moment are not of our concern.

Once the existence of the Kohn-Sham system is ensured by vLRGl,
one has to find a viable way to calculate the effective potential $v_s$.
Looking back at the inputs of the original problem,
$v_s(t)$ can be regarded as a functional of $\ket{\Psi_0}$, $\ket{\Phi_0}$ an $\vext$.
This will be denoted by $v_s(t)=F_t[\vext]$, in which the explicit dependence on 
initial states is omitted but always understood.
If we were given this functional, then the problem (\ref{KSeom}-\ref{KSiv})
would have been well defined and admit, like the problem (\ref{MBeom}-\ref{MBiv}),
one and only one solution.

A \textit{density}-functional formulation of the problem, which we are allowed to do thanks to vLRGl,
implies the definition of quantities that depend on the density and not on the external potential.
We cannot use directly $v_s(t)=\tilde{F}_t[n]$, even though vLRGl ensures its existence and uniqueness,
since the problem (\ref{KSeom}-\ref{KSiv}) would have no information at all
about the external potential that drives the evolution of (\ref{MBeom}).

To feed in some information about the external potential,
one usually \cite{rungegross} writes 
\be\label{vhxcdef}
v_s(t)=\vext(t)+\vhxc(t)
\ee
which in fact defines a new quantity, $\vhxc(t)$.
As long as vLRGl holds, we can write $\vhxc(t)=G_t[n]$, 
which expresses the functional dependence of $\vhxc(t)$ on $n$.

It should be emphasized that vLRGl states that,
given \textit{ the entire set} $\{n(t)|\;t\in[0,+\infty)\}$,
\textit{the entire set} $\{\vhxc(t)|\;t\in[0,+\infty)\}$ is unambiguously identified;
however, it does not specify how single elements of these sets are linked.
More specifically, it does not tell whether information on future densities is
necessary to construct $\vhxc(t)$ or not.
As discussed in \cite{sch,neepacausality}, it appears that this might threaten the predictivity of the approach.
For the standard electronic many-body problem of \cite{rungegross},
an argument to prove that these worries are unjustified was presented in \cite{neepacausality},
in which it was explained how $\vhxc(t)$ requires no information on the future to be completely determined.
However, this argument is specific for the Hamiltonian and the choice of densities of \cite{rungegross}.
Moreover, part of it is in conflict with the initial assumptions of \cite{rungegross},
as I shall soon explain.
In the following section I shall present a simpler model, which will allows me to enlighten
some limits of the original argument of \cite{neepacausality} and suggest possible solutions.

%%%%%%%%%%%%%%%%%%%%%%%%%%%%%%%%%%%%%%%%%%%%%%%%%%%%%%%%%%%%%%%%%%%%%%%%%%%%%%%%%%%%%%%%
%%%%        %%%%        %%%%        %%%%        %%%%        %%%%        %%%%        %%%%        %%%%        %%%%        %%%%        
%%%%%%%%%%%%%%%%%%%%%%%%%%%%%%%%%%%%%%%%%%%%%%%%%%%%%%%%%%%%%%%%%%%%%%%%%%%%%%%%%%%%%%%%

\section{The case of the Hubbard Dimer}\label{Shubbard}

%%%%%%%%%%%%%%%%%%%%%%%%%%%%%%%%%%%%%%%%%%%%%%%%%%%%%%%%%%%%%%%%%%%%%%%%%%%%%%%%%%%%%%%%
\subsection{Definition of the model and corresponding Kohn-Sham system}

I now consider the Hubbard dimer in presence of an external field.
Such a model has been analyzed in a TDDFT framework in several works
\cite{fuks2013,fuks2014a,fuks2014b}
but its relevance to the realistic electronic many-body problem is not of our concern,
as I use it only as prototype of a class of problems for which
the argument of \cite{neepacausality} does not apply.
The model is defined by
\be\label{MBH}
\hat{H}\equiv-\tau \sum_{\sigma=\uparrow,\downarrow}\left(
	\hat{c}^\dagger_{L\sigma}\hat{c}_{R\sigma} 
		+ \hat{c}^\dagger_{R\sigma}\hat{c}_{L\sigma} \right)+
	U\left(\hat{n}_{L\uparrow}\hat{n}_{L\downarrow}+
		\hat{n}_{R\uparrow}\hat{n}_{R\downarrow}\right)+
	v(t)\frac{\left( \hat{n}_L-\hat{n}_R\right)}{2}
\ee
%%1of2
%\begin{multline}\label{MBH}
%\hat{H}\equiv-\tau \sum_{\sigma=\uparrow,\downarrow}\left(
%	\hat{c}^\dagger_{L\sigma}\hat{c}_{R\sigma} 
%		+ \hat{c}^\dagger_{R\sigma}\hat{c}_{L\sigma} \right)+\\
%	+U\left(\hat{n}_{L\uparrow}\hat{n}_{L\downarrow}+
%		\hat{n}_{R\uparrow}\hat{n}_{R\downarrow}\right)+\\
%	+ v(t)\frac{\left( \hat{n}_L-\hat{n}_R\right)}{2}
%\end{multline}
where $\hat{n}_i\equiv\sum_\sigma \hat{c}^\dagger_{i\sigma}\hat{c}_{i\sigma}$,
 and $\{\hat{c}_{i\sigma},\hat{c}^\dagger_{j\rho}\}= \delta_{ij} \delta_{\sigma\rho}$,
with $i,j=R,L$ and $\sigma,\rho=\uparrow,\downarrow$, all other anticommutators being zero,
and by a given initial state $\ket{\Psi(t=0)}=\ket{\Psi_0}$.

If one is interested only in the value of $\braket{\hat{n}_L}$ and $\braket{\hat{n}_R}$,
one can attempt to solve the problem by means of the approach above outlined.
In fact, the value of $\braket{\hat{n}_L+\hat{n}_R}$ is only determined by the initial state, 
since $[\hat{n}_L+\hat{n}_R,H(t)]=0$.
We can then concentrate on the observable $n(t)$,
where
\be
\hat{n}\equiv\frac{\hat{n}_L-\hat{n}_R}{2}.
\ee
Following \cite{fuks2014b}, I chose the corresponding Kohn-Sham 
to be characterized by the following Hamiltonian:
\be\label{KSH}
\hat{H}_s\equiv-\tau \sum_{\sigma=\uparrow,\downarrow}\left(
	\hat{c}^\dagger_{L\sigma}\hat{c}_{R\sigma} 
		+ \hat{c}^\dagger_{R\sigma}\hat{c}_{L\sigma} \right)
	+  v_s(t)\hat{n}.
\ee
For simplicity, I consider $\ket{\Phi(t=0)}=\ket{\Psi_0}$.
Then, following the procedure outlined in the previous section, I define
\be\label{vhxcdef1}
  \vhxc(t)\equiv   v_s(t)-  v(t).
\ee
Such a function is determined by the condition
\be\label{n=n}
  n_{MB}(t)=  n_{KS}(t).
\ee
This condition can be recast as
\be
\left\{
\begin{array}{c}

			   n_{KS}(0)=  n_{MB}(0)\\
		\dot {n}_{KS}(0)=\dot {n}_{MB}(0)\\
		\ddot {n}_{KS}(t)=\ddot {n}_{MB}(t),
\end{array}
\right.
\ee
The first two identities are satisfied no matter what effective potential is chosen,
as long as the two initial states are taken to be the same.
So we can say that $\vhxc(t)$ is in fact defined by the condition 
\be\label{ddot}
\ddot {n}_{KS}(t)=\ddot {n}_{MB}(t).
\ee
The quantity $\ddot {n}(t)$ can be connected with the contemporary 
external potential by means of the equations of motion (\ref{KSeom}) and (\ref{MBeom}):
\be\label{MBlfe}
\vext(t)\braket{\hat{T}}_{MB}=\ddot{n}(t)+4 \tau^2 n(t)+\braket{[[\hat{n},\hat{T}],\hat{W}]}_{MB}.
\ee
Similarly, for the KS system we have:
\be\label{KSlfe}
v_s(t)\braket{\hat{T}}_{KS}=\ddot{n}(t)+4 \tau^2 n(t).
\ee
In fact, using (\ref{ddot},\ref{MBlfe},\ref{KSlfe},\ref{vhxcdef1})
it is possible to derive that
\be\label{vhxcdef2}
  \vhxc(t)=  v(t)\left(\frac{\braket{\hat{T}}_{MB}}{\braket{\hat{T}}_{KS}}-1\right)
-\frac{\braket{[[\hat{n},\hat{T}],\hat{W}]}_{MB}}{\braket{\hat{T}}_{KS}}.
\ee
Vice versa, as long as (\ref{vhxcdef2}) holds, equation (\ref{ddot}) and hence (\ref{n=n}) follow.

%%%%%%%%%%%%%%%%%%%%%%%%%%%%%%%%%%%%%%%%%%%%%%%%%%%%%%%%%%%%%%%%%%%%%%%%%%%%%%%%%%%%%%%%
\subsection{Taylor-expandability and Causality}

Adapting the constructive algorithm of \cite{vL1999},
it is possible to prove that a set of $\vhxc$ satisfying  (\ref{vhxcdef2}) does exist,
at least if we restrict our attention to the set of external potentials
that are Taylor-expandable in the time variable (`Te' potentials, from now on).
In fact, under this restriction, adapting the entire argument of \cite{vL1999} 
allows to prove the vLRGl for this case.
This also allows us to density-functionalize the theory.
We can then correctly regard $\vhxc(t)$ as a functional of the density only.
I now want to investigate the property of `predictivity' of the approach, 
in the sense of uniqueness of the solution of the KS system (\ref{KSeom},\ref{KSiv}),
with $\vhxc(t)=G_t[n]$.

Following \cite{neepacausality}, one might start by wondering about the dependence of $\vhxc(t)$
on preceding ($0\leq t'<t$), contemporary ($t'=t$), and successive ($t'>t$) densities $n(t')$.
In \cite{neepacausality} it was argued that, for any $T>0$,
$\vhxc(t)$ with $0<t<T$ cannot depend on densities $n(t)$ with $t>T$,
using the Runge-Gross theorem and the fact that two external potentials that are the same in $[0,T]$
but differ for $t>T$ must give different densities at times $t>T$.
However, for Te external potentials this very first step is problematic,
for two Te functions equal in a finite interval $0<\epsilon_1<t<\epsilon_2<T$ are in fact equal
everywhere within the radius of convergence of the Taylor series.
This means that, when Taylor expandable external potentials are considered, 
as we did in order to prove the vLRGl,
two external potentials that were the same in a finite interval in the past
will be necessarily the same also in the future (up to the radius of convergence, i.e. until vLRGl holds).
Equivalently, within the radius of convergence of the Taylor expansion of the density,
any dependence on the density on a finite interval in the past can be recast as 
a dependence on the density on a finite interval in the future, and vice versa.
In this case, the intuitive notion of $\vhxc(t)$ being `causal', 
in the sense that does not depend on future densities,
becomes meaningless. 

Nonetheless, a meaningful concept of `causality' is not necessary to discuss
the predictivity of the approach, that remains a non-trivial problem.
In order to proceed, we look back at equation (\ref{vhxcdef2})
and consider separately the objects in the form of $\braket{\Op}$ from $v(t)$.

%%%%%%%%%%%%%%%%%%%%%%%%%%%%%%%%%%%%%%%%%%%%%%%%%%%%%%%%%%%%%%%%%%%%%%%%%%%%%%%%%%%%%%%%
\subsection{Density Functionalization (1of2)}

The contemporary expectation values $\braket{\Op}$ in (\ref{vhxcdef2})
are unambiguously identified by the state vectors $\ket{\Psi(t)}$
and $\ket{\Phi(t)}$. 
In order to argue that these objects, when regarded as functionals of the density only,
pose no threat to the time propagation of the Kohn-Sham system,
we extend the second part of the argument of \cite{neepacausality},
based on a propagation on a finite time-grid.

More specifically, we consider a discrete time variable and a simple rule for translating
a differentiation on a continuous variable, namely
\be\label{rule}
\frac{df(t)}{dt}\rightarrow \frac{f_{i+1}-f_{i}}{\Delta}.
\ee
To compare results on the time-grid with the continuous limit,
we consider only terms of leading order in $\Delta$, 
since the limit $\Delta\rightarrow 0$ makes (\ref{rule}) an identity.

If we denote the many-body and Kohn-Sham state
at the time step $i$ with $\psi_i$ and $\phi_i$, respectively, we can write
\bea
\psi_{i+1}&=&F(v_i,\psi_i)\\
\phi_{i+1}&=&G(v^s_i,\phi_i)\label{phi}\\
v_i&=&V(n_{i+2},n_{i+1},\psi_i,\phi_i)\\
v_i^s&=&V^s(v_i,\psi_i,\phi_i)\label{vs}\\
n_{i+1}&=&N(n_i,J_i)
\eea
with $F(.)$ obtained from (\ref{MBeom},\ref{MBH},\ref{rule}), 
 $G(.)$ from (\ref{KSeom},\ref{KSH},\ref{rule}), 
 $V(.)$ from (\ref{MBlfe},\ref{rule}),%
\footnote{In fact, an entry in $V(.)$ for $\phi_i$ is redundant, since $v_i$
is only determined by $n_{i+2}$, $n_{i+1}$ and $\psi_i$. However,
this is specific for the model here considered,
while in general also $\phi_i$ may be required.}
 $V^s(.)$ from (\ref{vhxcdef1},\ref{vhxcdef2}),
and $N(.)$ from the continuity equation
\be\label{cont}
\dot {n}(t)=-\bra{\Phi(t)} \hat{J}\ket{\Phi(t)},
\ee
with $\hat{J}\equiv i[\hat{n},\hat{T}]$.
Then, for the two terms appearing in (\ref{vhxcdef2}) we can write 
\be\label{OO}
\frac{\braket{\Op_1}_{MB}}{\braket{\Op_2}_{KS}}\rightarrow L(\psi_i,\phi_i)
\ee
for which
\begin{widetext}
\bea
L(\psi_i,\phi_i)&=&L(F(v_{i-1},\psi_{i-1}),G(v^s_{i-1},\psi_{i-1}))=\\
			  &=&L\left(F\left(V\left(n_{i+1},n_{i},\psi_{i-1},\phi_{i-1}\right),\psi_{i-1}\right),
	G\left(V^s(V(n_{i+1},n_{i},\psi_{i-1},\phi_{i-1}),\psi_{i-1},\phi_{i-1}),\psi_{i-1}\right)\right)=\\
	&\equiv&M(n_{i+1},n_{i},\psi_{i-1},\phi_{i-1})=\\
	&=&M(N(n_i,J_i),n_{i},\psi_{i-1},\phi_{i-1})=\\
	&\equiv&L_{(1)}(J_i,n_{i},\psi_{i-1},\phi_{i-1})=\\
    &=&L_{(1)}(J_i,n_{i},F(v_{i-2},\psi_{i-2}),G(v^s_{i-2},\psi_{i-2}))=\\
    &=&L_{(1)}(J_i,n_{i},F\left(V\left(n_{i},n_{i-1},\psi_{i-2},\phi_{i-2}\right),\psi_{i-2}\right),
	G\left(V^s(V(n_{i},n_{i-1},\psi_{i-2},\phi_{i-2}),\psi_{i-2},\phi_{i-2}),\psi_{i-2}\right))=\\
    &\equiv&L_{(2)}(J_i,n_{i},n_{i-1},\psi_{i-2},\phi_{i-2})=    \\
		&&\;\;\;\;\; \vdots \\
	&=&L_{(i)}(J_i,n_{i},n_{i-1},...,n_1,\psi_{0},\phi_{0}).\label{L}
\eea
\end{widetext}
This means that on a time-grid these terms are unambiguously determined by:
the Kohn-Sham state vector at same time step (which enters through $J_i$), 
the set of contemporary and previous densities, and the initial states.
Since at a given time step, these ingredients are in principle all at our disposal,
they pose no threat to the time-propagation of such a discretized KS system.
While rigorous for the discretized problem, this argument is only a first step towards
a complete proof for the continuous time problem.

Nonetheless, when applied to the electronic problem of \cite{rungegross},
the argument supports the claim of \cite{neepacausality}
that terms in the form of (\ref{OO}) pose no threat to the time-propagation of the 
corresponding KS system.

In fact, following farther the argument, we are also led to conclude that
the $\vhxc({\bf r},t)$ of \cite{rungegross} cannot be
determined by sole knowledge of initial states, contemporary and past densities,
but also requires knowledge of the divergence of the contemporary Kohn-Sham current
${\bf \nabla}\cdot {\bf j}(t)$, which appears in the equivalent of (\ref{cont}).
The importance of such a term was recognized in \cite{neepacausality},
but only at $t=0$, when its value can be obtained from the initial KS state.
Pushing farther their own argument, as done above, suggests that
the dependence of $\vhxc$ on ${\bf \nabla}\cdot {\bf j}$ for finite times,
rather than remaining a dependence on the initial ${\bf \nabla}\cdot {\bf j}(t=0)$,
it becomes a dependence on the \textit{contemporary} ${\bf \nabla}\cdot {\bf j}$.

Such a dependence is not incompatible with vLRGl, 
for which we expect that complete knowledge of the density suffices to identify the system. 
$J_i$ has indeed been introduced to encode the information contained in $n_{i+1}$,
which, in the language of the continuous time case, translates into $\dot{n}(t)$.
Choosing of expressing such an information in terms of $J(t)$, which one can express 
in terms of $\ket{\Phi(t)}$, rather than $\dot{n}(t)$,
avoids the complications of having a time derivative of the state
on the right-hand side of the equation of motion (\ref{KSeom}).

%%%%%%%%%%%%%%%%%%%%%%%%%%%%%%%%%%%%%%%%%%%%%%%%%%%%%%%%%%%%%%%%%%%%%%%%%%%%%%%%%%%%%%%%
\subsection{Density Functionalization (2of2)}\label{Sdft2}

Once the $n$ dependence of the terms $\braket{\Op}$ has been clarified,
to some extend at least, we can go back to (\ref{vhxcdef2}) and discuss the presence of $v(t)$.
This needed to not to be discussed in \cite{neepacausality}, because of no explicit $v(t)$
dependence in the corresponding equation.
More precisely, in the standard TDDFT case, the equivalent of (\ref{MBlfe}) and (\ref{KSlfe}) reads
\be
{\bf \nabla}\cdot[n({\bf r},t){\bf \nabla}v({\bf r},t)]+\braket{\Op_1}_{MB}
\ee
and
\be
{\bf \nabla}\cdot[n({\bf r},t){\bf \nabla}v_s({\bf r},t)]+\braket{\Op_2}_{MB}
\ee
with $\Op_2$ and $\Op_1$ two appropriate operators,
$n({\bf r},t)$ the electronic charge-density,
$v({\bf r},t)$ the external field and 
$v_s({\bf r},t)$ the effective field of the corresponding KS system \cite{vL1999}.
Since both potentials are multiplied by the electronic charge-density,
a quantity that is the same for the many-body and the KS system, the explicit $v(t)$-dependence
in the corresponding $\vhxc({\bf r},t)$ cancels out.

The fact that in general $\braket{\hat{T}}_{MB}/\braket{\hat{T}}_{KS}$
does not simplify to $1$, leading in our case to the missed cancellation,
can be proved by considering the specific case of
$\tau=U=1$, $v(t)=\sin (t)$, and $\ket{\Psi_0}$ being the half-filled
ground-state of the Hamiltonian at $t=0$, for which
\be
\begin{array}{l}
\braket{\hat{T}}_{MB}=-2+t^2-\frac{7}{6}t^4+\frac{179}{360}t^6+\mathcal{O}(t^8)\\
\braket{\hat{T}}_{KS}=-2+\frac{1}{4}t^4+\frac{31}{72}t^6+\mathcal{O}(t^8),
\end{array}
\ee
as one can prove by using the constructive algorithm of \cite{vL1999}.

The expression on the right-hand of (\ref{vhxcdef2}) side is therefore not `universal',
in the sense that it explicitly depends on the system via $v(t)$.
However, this dependence can be removed if we 
use again equation (\ref{MBlfe}) to extract the information
about the density $n$ contained in $v(t)$. This leads to
\be\label{vddot}
  v(t)=F_t[n]+G_t[n] \ddot{n}.
\ee
with $F$ and $G$ some suitable functionals.
While for these two functionals the argument of the previous section applies,
the presence of $\ddot{n}(t)$ needs a dedicated analysis.

Even though one could discuss the consequences of the presence of $\ddot {n}(t)$
on the time-grid, in fact we do not need to, as they can be cleared enough by 
directly looking at the original problem in continuous time.
When (\ref{vddot}) is plugged back into the equation of motion (\ref{KSeom}),
an explicit dependence on the second-time derivative of $\ket{\Phi(t)}$ is introduced.
Even assuming that the conclusions of the previous sections 
can be safely generalised to the continuous time case,
the presence of a double time-derivative of $\ket{\Phi(t)}$ changes 
the status of the equation (\ref{KSeom}) from first to second-order.
This implies that the initial value (\ref{KSiv}) is no longer sufficient 
to ensure one single solution to the problem.
In other words,
the density-functionalization of the Kohn-Sham system (\ref{KSH}) leads to a problem that,
despite being an actual exact reformulation of the many-body problem,
does not admit only one solution,
and hence it fails to unambiguously characterize the solution of the initial many-body problem,
even if all functionals were known.
Using the terminology of \cite{sch,neepacausality} one could say
that this Kohn-Sham system \textit{is not predictive}.

%%%%%%%%%%%%%%%%%%%%%%%%%%%%%%%%%%%%%%%%%%%%%%%%%%%%%%%%%%%%%%%%%%%%%%%%%%%%%%%%%%%%%%%%

\subsection{Restoring Predictivity}

Applying the recipe of Section \ref{Sprocedure} to the Hubbard dimer above defined led
to a reformulation of the original problem that, despite being exact, cannot be used to
calculate the wanted densities.

This problem can in fact be avoided if we use a slightly modified recipe,
which however has the cost of introducing a new (unknown) functional of the density.
More specifically, the way information about $\vext(t)$ is extracted from $v_s(t)$ has to be modified
as follows:
\be\label{ffdef}
\left\{\begin{array}{l}
v_s(t)= v_1(t) \vext(t)+ v_2(t)\\
v_1(t)\equiv \frac{\braket{\hat{T}}_{MB}}{\braket{\hat{T}}_{KS}}\\
v_2(t)\equiv -\frac{\braket{[[\hat{n},\hat{T}],\hat{W}]}_{MB}}{\braket{\hat{T}}_{KS}}.
\end{array}\right.
\ee
When $v_1$ and $v_2$ are regarded as functionals of $n$,
the argument used above for $\braket{\Op}$ applies again and the problems cause by
the presence of $\ddot{n}(t)$ are avoided.

%%%%%%%%%%%%%%%%%%%%%%%%%%%%%%%%%%%%%%%%%%%%%%%%%%%%%%%%%%%%%%%%%%%%%%%%%%%%%%%%%%%%%%%%
%%%%        %%%%        %%%%        %%%%        %%%%        %%%%        %%%%        %%%%        %%%%        %%%%        %%%%        
%%%%%%%%%%%%%%%%%%%%%%%%%%%%%%%%%%%%%%%%%%%%%%%%%%%%%%%%%%%%%%%%%%%%%%%%%%%%%%%%%%%%%%%%

\section{The Density-Functional Time-Dependent Kohn-Sham Approach (2 of 2)}\label{Sprocedure2}
%%%%%%%%%%%%%%%%%%%%%%%%%%%%%%%%%%%%%%%%%%%%%%%%%%%%%%%%%%%%%%%%%%%%%%%%%%%%%%%%%%%%%%%%
\subsection{A New Recipe}

The case of the Hubbard dimer here considered is not an isolated exception, 
as we shall see in the next section.
In fact we can use what learned from this example to identify a \textit{class} of problems 
that, if approached in the standard way, present the same complications.

Given an Hamiltonian and a certain choice of densities for which the vLRGl holds,
we can connect the external potential(s) to the effective one(s) 
by differentiating the densities and using the equations of motion (\ref{MBeom}-\ref{KSeom}).
This would lead to an expression that in the simplified notation of Section \ref{Sprocedure} would read
\be\label{testeq}
\braket{\Op_1[v_s(t)] }_{KS} +\braket{\Op_2}_{KS}=
\braket{\Op_3[\vext (t)]}_{MB} +\braket{\Op_4}_{MB}
\ee
with $\Op_i[f(t)]$ acting as an operator on the Hilbert space of state vectors and possibly
a differential operator on the argument $f(t)$. 
If
\be
\braket{\Op_1[f(t)] }_{KS} = \braket{\Op_3[f (t)]}_{MB},
\ee
then we can say that the problem (Hamiltonian$+$choice of densities)
is \textit{RG-like} and the KS system is as predictive as the one of standard TDDFT \cite{rungegross};
otherwise, we can say that it is \textit{RG-unlike} and
considering the Hartree-exchange-correlation potential defined by $\vhxc (t) \equiv v_s (t) - \vext(t)$
as a functional of the density only leads to a reformulation of the many-body problem
that does not admit one unique solution, making the density-functionalized Kohn-Sham system not predictive.

In case the operator $\Op_i[f(t)]$ is purely multiplicative on its argument: 
$\Op_i[f(t)]\rightarrow \Op_i f(t)$, like in our example of the Hubbard dimer, 
it is possible to fix this problem by simply considering
\be
\left\{\begin{array}{l}
v_s(t)= v_1(t) \vext(t)+ v_2(t)\\
v_1(t)\equiv \frac{\braket{\Op_3}_{MB}}{\braket{\Op_1}_{KS}}\\
v_2(t)\equiv \frac{\braket{\Op_4}_{MB}-\braket{\Op_2}_{KS}}{\braket{\Op_1}_{KS}}
\end{array}\right.
\ee
and regarding $v_1(t)$ and $v_2(t)$ as functionals of the density separately.

When the density(/ies) is(/are) chosen to be the 
expectation value of the local operator(s) conjugate to the external potential(s)
in the many-body Hamiltonian, one has
\bea
&&\braket{\Op_1[v_s(t)]}=\braket{[\hat{V}_s,[\hat{T},\hat{n}]]}\\
&&\braket{\Op_2}=\braket{[\hat{T},[\hat{T},\hat{n}]]}\\
&&\braket{\Op_3[\vext (t)]}=\braket{[\hat{V},[\hat{T}+\hat{W},\hat{n}]]}\\
&&\braket{\Op_4}=\braket{[\hat{T}+\hat{W},[\hat{T}+\hat{W},\hat{n}]]}.
\eea
by virtue of $[\hat{V},\hat{n}]=0$.

%%%%%%%%%%%%%%%%%%%%%%%%%%%%%%%%%%%%%%%%%%%%%%%%%%%%%%%%%%%%%%%%%%%%%%%%%%%%%%%%%%%%%%%%
\subsection{Relativistic Electrons}

Another example of the class of \textit{RG-unlike} problems 
is provided by the Kohn-Sham approach to QED defined in \cite{rmb}, 
to which I refer the reader for the notation of this section.
In that work the four-potential $J^\mu$ is chosen as `density'. 
As one can easily verify from equation (31) of \cite{rmb},
our equation (\ref{testeq}) becomes
\be
\braket{\hat{\Op}^{\mu\nu}}_{KS}a^{s}_\nu+\braket{\hat{\mathcal{Q}}^\mu_{\rm kin}}_{KS}
=\braket{\hat{\Op}^{\mu\nu}}_{QED}a^{\rm ext}_\nu
+\braket{\hat{\mathcal{Q}}^\mu_{\rm kin}+\hat{\mathcal{Q}}^\mu_{\rm int}}_{QED}
\ee
%%2of2
%\begin{multline}
%\braket{\hat{\Op}^{\mu\nu}}_{KS}a^{s}_\nu+\braket{\hat{\mathcal{Q}}^\mu_{\rm kin}}_{KS}=\\
%=\braket{\hat{\Op}^{\mu\nu}}_{QED}a^{\rm ext}_\nu
%+\braket{\hat{\mathcal{Q}}^\mu_{\rm kin}+\hat{\mathcal{Q}}^\mu_{\rm int}}_{QED}
%\end{multline}
with 
\be
\hat{\Op}^{\mu\nu}\equiv \hat{\psib}\left(\gamma^\mu \gamma^0 \gamma^\nu
-\gamma^\nu \gamma^0 \gamma^\mu \right)\hat{\psi} 
\ee
where $\hat{\psi}$ is the Dirac field operator, $\psib\equiv \psi^\dagger\gamma^0$
and $\gamma^\mu$ the gamma matrices acting on the spin degrees of freedom of $\psi$,
and in which explicit spacetime dependence has been omitted.
Since $\braket{\hat{\Op}^{\mu\nu}}_{KS}\neq \braket{\hat{\Op}^{\mu\nu}}_{MB}$
the approach has the problem of predictivity here discussed, if $a^\mu_{\rm Hxc}(x)$
is regarded as a functional of the four-current only.\footnote{%
In applying the argument of Section \ref{Sdft2} one might think
that using again equation (31) of \cite{rmb} to connect $a^\mu_{\rm ext}(x)$ to $\dot{j}^\mu$,
the resulting Dirac-Kohn-Sham equation would not become of second order,
but remaining of first. 
If so, the argument on the number of solutions would no longer apply.
However, only two components of $a^\mu_{\rm ext}(x)$ are determined by (31),
since the kernel of the matrix $\braket{\Op^{\mu\nu}}$ is of dimension 2.
Even if one component is fixed by a gauge condition,
another component remains to fix and for this one should necessarily
look at higher derivatives.}

The case of QED is particularly instructive because it shows that the predictivity issue
here discussed does not seem to be necessarily inherent the many-body Hamiltonian considered,
for the choice of densities and the Kohn-Sham system can also be crucial.
In \cite{rugge2} a different set of densities was chosen for the same QED problem,
namely the four quantities $\psi^\dagger \gamma^\mu \psi\equiv P^\mu$.
In this case, the equivalent of (\ref{testeq}) calculated in the Coulomb gauge
(using equations (63) and (65) of \cite{rugge2}) leads to 
$ \braket{\hat{\Op}_1[v_s(t)]}_{KS} \rightarrow P^0 \vec{a}_s $ and
$ \braket{\hat{\Op}_3[\vext(t)]}_{MB} \rightarrow P^0 \vec{a}_{\rm ext} $.
Being $P^0$ one of the chosen densities, the identity
$ \braket{\hat{\Op}_1[\vext (t)]}_{KS} = \braket{\hat{\Op}_3[\vext(t)]}_{MB} $
is in fact fulfilled making the system of \cite{rugge2} \textit{RG-like}.

\section*{Conclusions}\label{Sconclusions}

In this work the arguments used in \cite{neepacausality}
to support the predictive character of the Kohn-Sham system as an initial-value problem
within a TDDFT framework were reviewed and extended.

It was pointed out that the original argument cannot be applied if external potentials 
that are Taylor expandable in time are considered. 
Then, in order to study the `predictivity' of a density-functionalized Kohn-Sham system,
identified with the property of such a system of admitting one 
and only one solution when regarded as an initial-value problem,
an argument on a time-grid of infinitesimal pace has been developed.
When applied to the standard TDDFT case, such an argument, on one hand, 
confirms the claims of \cite{neepacausality} about predictivity of standard TDDFT; 
on the other hand, it suggests that, 
in order to preserve the differential structure of the Kohn-Sham equation
and have a predictive Kohn-Sham system, $\vhxc({\bf r},t)$ must be regarded 
as functional of contemporary and previous densities,
the many-body and Kohn-Sham initial state and \textit{the divergence of the contemporary Kohn-Sham current.}
The argument is however rigorous only on a time-grid and the
problem demands for further investigations.

The specificity of the original argument of \cite{neepacausality} to the TDDFT problem
has also been enlightened by showing another many-body problem, a time-dependent Hubbard dimer,
whose Kohn-Sham system is in fact not predictive,
as long as the corresponding $\vhxc(t)\equiv v_s(t)-\vext(t)$ 
is regarded as a functional of the density only.
It was proved that the Kohn-Sham equation resulting from a density-functionalization of $\vhxc$
was requiring more boundary-conditions then provided, failing to being characterized by a unique solution,
and hence failing to make a `predictive' system.
The origin of such a problem was identified, allowing to derive a criteria
to fulfill for a Kohn-Sham system for not being affected by the same predictivity issue.

In some simple cases, like for the Hubbard dimer here considered, predictivity can be restored
by modifying the definition of the quantities to be considered functionals of the density.
More specifically, it was argued that, if the effective potential is decomposed as 
$v_s(t)=v_1(t) \vext(t)+v_2(t)$,
with $v_1(t)$ and $v_2(t)$ unambiguously defined as expectation values of some specific operators,
and the two potentials $v_1(t)$ and $v_2(t)$ are regarded as functionals of the density,
the Kohn-Sham initial-value problem has in fact only one solution.

\section*{Acknowledgements}

The research leading to these results has received funding from the
European Research Council under the European Union's Seventh Framework Programme 
(FP/2007-2013) / ERC grant agreement no. 320971.

\bibliography{bibliography}

\begin{thebibliography}{11}%
\makeatletter
\providecommand \@ifxundefined [1]{%
 \@ifx{#1\undefined}
}%
\providecommand \@ifnum [1]{%
 \ifnum #1\expandafter \@firstoftwo
 \else \expandafter \@secondoftwo
 \fi
}%
\providecommand \@ifx [1]{%
 \ifx #1\expandafter \@firstoftwo
 \else \expandafter \@secondoftwo
 \fi
}%
\providecommand \natexlab [1]{#1}%
\providecommand \enquote  [1]{``#1''}%
\providecommand \bibnamefont  [1]{#1}%
\providecommand \bibfnamefont [1]{#1}%
\providecommand \citenamefont [1]{#1}%
\providecommand \href@noop [0]{\@secondoftwo}%
\providecommand \href [0]{\begingroup \@sanitize@url \@href}%
\providecommand \@href[1]{\@@startlink{#1}\@@href}%
\providecommand \@@href[1]{\endgroup#1\@@endlink}%
\providecommand \@sanitize@url [0]{\catcode `\\12\catcode `\$12\catcode
  `\&12\catcode `\#12\catcode `\^12\catcode `\_12\catcode `\%12\relax}%
\providecommand \@@startlink[1]{}%
\providecommand \@@endlink[0]{}%
\providecommand \url  [0]{\begingroup\@sanitize@url \@url }%
\providecommand \@url [1]{\endgroup\@href {#1}{\urlprefix }}%
\providecommand \urlprefix  [0]{URL }%
\providecommand \Eprint [0]{\href }%
\providecommand \doibase [0]{http://dx.doi.org/}%
\providecommand \selectlanguage [0]{\@gobble}%
\providecommand \bibinfo  [0]{\@secondoftwo}%
\providecommand \bibfield  [0]{\@secondoftwo}%
\providecommand \translation [1]{[#1]}%
\providecommand \BibitemOpen [0]{}%
\providecommand \bibitemStop [0]{}%
\providecommand \bibitemNoStop [0]{.\EOS\space}%
\providecommand \EOS [0]{\spacefactor3000\relax}%
\providecommand \BibitemShut  [1]{\csname bibitem#1\endcsname}%
\let\auto@bib@innerbib\@empty
%</preamble>
\bibitem [{\citenamefont {Maitra}\ \emph {et~al.}(2008)\citenamefont {Maitra},
  \citenamefont {van Leeuwen},\ and\ \citenamefont {Burke}}]{neepacausality}%
  \BibitemOpen
  \bibfield  {author} {\bibinfo {author} {\bibfnamefont {N.~T.}\ \bibnamefont
  {Maitra}}, \bibinfo {author} {\bibfnamefont {R.}~\bibnamefont {van Leeuwen}},
  \ and\ \bibinfo {author} {\bibfnamefont {K.}~\bibnamefont {Burke}},\ }\href
  {\doibase 10.1103/PhysRevA.78.056501} {\bibfield  {journal} {\bibinfo
  {journal} {Phys. Rev. A}\ }\textbf {\bibinfo {volume} {78}},\ \bibinfo
  {pages} {056501} (\bibinfo {year} {2008})}\BibitemShut {NoStop}%
\bibitem [{\citenamefont {Runge}\ and\ \citenamefont
  {Gross}(1984)}]{rungegross}%
  \BibitemOpen
  \bibfield  {author} {\bibinfo {author} {\bibfnamefont {E.}~\bibnamefont
  {Runge}}\ and\ \bibinfo {author} {\bibfnamefont {E.~K.~U.}\ \bibnamefont
  {Gross}},\ }\href {\doibase 10.1103/PhysRevLett.52.997} {\bibfield  {journal}
  {\bibinfo  {journal} {Phys. Rev. Lett.}\ }\textbf {\bibinfo {volume} {52}},\
  \bibinfo {pages} {997} (\bibinfo {year} {1984})}\BibitemShut {NoStop}%
\bibitem [{\citenamefont {van Leeuwen}(1999)}]{vL1999}%
  \BibitemOpen
  \bibfield  {author} {\bibinfo {author} {\bibfnamefont {R.}~\bibnamefont {van
  Leeuwen}},\ }\href {\doibase 10.1103/PhysRevLett.82.3863} {\bibfield
  {journal} {\bibinfo  {journal} {Phys. Rev. Lett.}\ }\textbf {\bibinfo
  {volume} {82}},\ \bibinfo {pages} {3863} (\bibinfo {year}
  {1999})}\BibitemShut {NoStop}%
\bibitem [{\citenamefont {Schirmer}\ and\ \citenamefont {Dreuw}(2007)}]{sch}%
  \BibitemOpen
  \bibfield  {author} {\bibinfo {author} {\bibfnamefont {J.}~\bibnamefont
  {Schirmer}}\ and\ \bibinfo {author} {\bibfnamefont {A.}~\bibnamefont
  {Dreuw}},\ }\href {\doibase 10.1103/PhysRevA.75.022513} {\bibfield  {journal}
  {\bibinfo  {journal} {Phys. Rev. A}\ }\textbf {\bibinfo {volume} {75}},\
  \bibinfo {pages} {022513} (\bibinfo {year} {2007})}\BibitemShut {NoStop}%
\bibitem [{\citenamefont {Ullrich}(2012)}]{ullrich}%
  \BibitemOpen
  \bibfield  {author} {\bibinfo {author} {\bibfnamefont {C.}~\bibnamefont
  {Ullrich}},\ }\href {http://books.google.ch/books?id=j\_4nodMFCa4C} {\emph
  {\bibinfo {title} {Time-Dependent Density-Functional Theory: Concepts and
  Applications}}},\ Oxford Graduate Texts\ (\bibinfo  {publisher} {OUP
  Oxford},\ \bibinfo {year} {2012})\BibitemShut {NoStop}%
\bibitem [{\citenamefont {Ruggenthaler}\ and\ \citenamefont {van
  Leeuwen}(2011)}]{fixed}%
  \BibitemOpen
  \bibfield  {author} {\bibinfo {author} {\bibfnamefont {M.}~\bibnamefont
  {Ruggenthaler}}\ and\ \bibinfo {author} {\bibfnamefont {R.}~\bibnamefont {van
  Leeuwen}},\ }\href {http://stacks.iop.org/0295-5075/95/i=1/a=13001}
  {\bibfield  {journal} {\bibinfo  {journal} {EPL (Europhysics Letters)}\
  }\textbf {\bibinfo {volume} {95}},\ \bibinfo {pages} {13001} (\bibinfo {year}
  {2011})}\BibitemShut {NoStop}%
\bibitem [{\citenamefont {Fuks}\ \emph {et~al.}(2013)\citenamefont {Fuks},
  \citenamefont {Farzanehpour}, \citenamefont {Tokatly}, \citenamefont {Appel},
  \citenamefont {Kurth},\ and\ \citenamefont {Rubio}}]{fuks2013}%
  \BibitemOpen
  \bibfield  {author} {\bibinfo {author} {\bibfnamefont {J.~I.}\ \bibnamefont
  {Fuks}}, \bibinfo {author} {\bibfnamefont {M.}~\bibnamefont {Farzanehpour}},
  \bibinfo {author} {\bibfnamefont {I.~V.}\ \bibnamefont {Tokatly}}, \bibinfo
  {author} {\bibfnamefont {H.}~\bibnamefont {Appel}}, \bibinfo {author}
  {\bibfnamefont {S.}~\bibnamefont {Kurth}}, \ and\ \bibinfo {author}
  {\bibfnamefont {A.}~\bibnamefont {Rubio}},\ }\href {\doibase
  10.1103/PhysRevA.88.062512} {\bibfield  {journal} {\bibinfo  {journal} {Phys.
  Rev. A}\ }\textbf {\bibinfo {volume} {88}},\ \bibinfo {pages} {062512}
  (\bibinfo {year} {2013})}\BibitemShut {NoStop}%
\bibitem [{\citenamefont {Fuks}\ and\ \citenamefont
  {Maitra}(2014{\natexlab{a}})}]{fuks2014a}%
  \BibitemOpen
  \bibfield  {author} {\bibinfo {author} {\bibfnamefont {J.~I.}\ \bibnamefont
  {Fuks}}\ and\ \bibinfo {author} {\bibfnamefont {N.~T.}\ \bibnamefont
  {Maitra}},\ }\href {\doibase 10.1103/PhysRevA.89.062502} {\bibfield
  {journal} {\bibinfo  {journal} {Phys. Rev. A}\ }\textbf {\bibinfo {volume}
  {89}},\ \bibinfo {pages} {062502} (\bibinfo {year}
  {2014}{\natexlab{a}})}\BibitemShut {NoStop}%
\bibitem [{\citenamefont {Fuks}\ and\ \citenamefont
  {Maitra}(2014{\natexlab{b}})}]{fuks2014b}%
  \BibitemOpen
  \bibfield  {author} {\bibinfo {author} {\bibfnamefont {J.~I.}\ \bibnamefont
  {Fuks}}\ and\ \bibinfo {author} {\bibfnamefont {N.~T.}\ \bibnamefont
  {Maitra}},\ }\href {\doibase 10.1039/C4CP00118D} {\bibfield  {journal}
  {\bibinfo  {journal} {Phys. Chem. Chem. Phys.}\ }\textbf {\bibinfo {volume}
  {16}},\ \bibinfo {pages} {14504} (\bibinfo {year}
  {2014}{\natexlab{b}})}\BibitemShut {NoStop}%
\bibitem [{\citenamefont {Ruggenthaler}\ \emph {et~al.}(2011)\citenamefont
  {Ruggenthaler}, \citenamefont {Mackenroth},\ and\ \citenamefont
  {Bauer}}]{rmb}%
  \BibitemOpen
  \bibfield  {author} {\bibinfo {author} {\bibfnamefont {M.}~\bibnamefont
  {Ruggenthaler}}, \bibinfo {author} {\bibfnamefont {F.}~\bibnamefont
  {Mackenroth}}, \ and\ \bibinfo {author} {\bibfnamefont {D.}~\bibnamefont
  {Bauer}},\ }\href {\doibase 10.1103/PhysRevA.84.042107} {\bibfield  {journal}
  {\bibinfo  {journal} {Phys. Rev. A}\ }\textbf {\bibinfo {volume} {84}},\
  \bibinfo {pages} {042107} (\bibinfo {year} {2011})}\BibitemShut {NoStop}%
\bibitem [{\citenamefont {Ruggenthaler}\ \emph {et~al.}(2014)\citenamefont
  {Ruggenthaler}, \citenamefont {Flick}, \citenamefont {Pellegrini},
  \citenamefont {Appel}, \citenamefont {Tokatly},\ and\ \citenamefont
  {Rubio}}]{rugge2}%
  \BibitemOpen
  \bibfield  {author} {\bibinfo {author} {\bibfnamefont {M.}~\bibnamefont
  {Ruggenthaler}}, \bibinfo {author} {\bibfnamefont {J.}~\bibnamefont {Flick}},
  \bibinfo {author} {\bibfnamefont {C.}~\bibnamefont {Pellegrini}}, \bibinfo
  {author} {\bibfnamefont {H.}~\bibnamefont {Appel}}, \bibinfo {author}
  {\bibfnamefont {I.~V.}\ \bibnamefont {Tokatly}}, \ and\ \bibinfo {author}
  {\bibfnamefont {A.}~\bibnamefont {Rubio}},\ }\href {\doibase
  10.1103/PhysRevA.90.012508} {\bibfield  {journal} {\bibinfo  {journal} {Phys.
  Rev. A}\ }\textbf {\bibinfo {volume} {90}},\ \bibinfo {pages} {012508}
  (\bibinfo {year} {2014})}\BibitemShut {NoStop}%
\end{thebibliography}%
\end{document}